\documentclass[review]{elsarticle}

\usepackage{lineno,hyperref}
\modulolinenumbers[5]

\journal{Journal of \LaTeX\ Templates}

\usepackage{amsmath}
\usepackage{algorithm}
\usepackage{algorithmicx}
\usepackage{tabularx} 
\usepackage{algorithm}
\usepackage{algorithmicx}
\usepackage{algpseudocode} 

\begin{document}

\begin{frontmatter}

\title{\bfseries PPLS: A Privacy-Preserving Location-Sharing Scheme in Vehicular Social Networks}


\author[mymainaddress]{Chang Xu}

\author[mymainaddress]{Xuan Xie}

\author[mymainaddress]{Liehuang Zhu\corref{mycorrespondingauthor}}
\cortext[mycorrespondingauthor]{Corresponding author}
\ead{liehuangz@bit.edu.cn}

\author[mymainaddress]{Kashif Sharif}

\author[mymainaddress]{Chuan Zhang}

\author[mysecondaryaddress]{Xiaojiang Du}

\author[mytertiaryaddress]{Mohsen Guizani}

\address[mymainaddress]{Beijing Engineering Research Center of Massive Language Information Processing and Cloud Computing 
Application,
School of Computer Science and Technology, Beijing Institute of Technology, Beijing, China}
\address[mysecondaryaddress]{Department of Computer and Information Sciences, Temple University, Philadelphia, USA}
\address[mytertiaryaddress]{Department of Electrical and Computer Engineering, University of Idaho, Moscow, Idaho, USA}




\begin{abstract}
The recent proliferation of mobile computing has given rise to vehicular social networks (VSNs) which use  the Internet of Vehicles and social networks as  the main design elements. As one of the most critical components in VSNs, location sharing plays an important role in  helping vehicles share information and strengthen their social bonds. This, however, may compromise vehicles' privacy, including location information and social relationship details. Some solutions have been proposed to address these challenges. However, none of them considers privacy of inter-user threshold distance, which effectively can be used to identify vehicles, their friends, and location information, by malicious or undesired elements of the system. In order to overcome this limitation, we propose a secure distance comparison protocol. Furthermore, we present a privacy-preserving location-sharing scheme in VSNs, namely PPLS, which  allows vehicles to build more complex access control policies. The safety of our scheme is validated by the security analysis, and experimental results demonstrate the efficiency of PPLS scheme.
\end{abstract}


\end{frontmatter}

\section{Introduction}

With the fast development \& deployment of mobile computing, vehicular ad hoc networks have
become   important data transmission platforms and greatly promote  the realization of Intelligent Transport System (ITS). Recently,   the application goal of vehicular ad hoc  networks \cite{JamilJUR17} transforms from simply improving the safety of road traffic and the efficiency of transportation to vehicular social networks (VSNs), which deeply integrates the Internet of Vehicles (IoVs) \cite{HaouariMSBG17} and social networks. Through VSNs, vehicles can experience more comprehensive services.


VSNs can provide various services \cite{SedjelmaciSA14,zhang2017inbar}, including location-based services (LBSs). In LBSs, geographical locations of vehicles are exploited to provide information and entertainment services, since the location of a vehicle usually represents its contextual information \cite{DBLP:journals/wicomm/HoangLNW13}. As millions of applications based on LBSs are available, vehicles can easily obtain information such as restaurants, hotels, etc. In fact, as a fundamental component of VSNs, LBSs have become increasingly popular and important.  
 
While enjoying the convenience of location-based services, the privacy threats should not be ignored \cite{DuC08}. Especially after some research work \cite{huang2014achieving,wu2014mobifish,wu2014security} revealing horrifying security and privacy issues which have caused serious public concerns. In LBSs, users are expected to update their real-time location information and share it for better services. However, disclosing the location information is dangerous, since an adversary can track an individual and infer his/her preferences. This threat becomes more serious in VSNs as vehicles' location can be correlated with their profiles \cite{DBLP:conf/interact/BarkhuusD03}.  Hence, it is essential to protect vehicles' location privacy \cite{ChenZDFLY12,ZhangXYD16,ZhangXZSD15} in VSNs. 
 
To address these problems, a series of  research works have  been performed. A MobiShare system was presented by Wei et al. \cite{Wei2012MobiShare}, allowing users to share location information flexibly. Inspired by  \cite{Wei2012MobiShare}, Shen et al. \cite{shen2016efficient,DBLP:conf/3pgcic/LiuLCLJ13} proposed a system called N-Mobishare. Li et al.  \cite{li2014mobishare} proposed MobiShare+ which reduces the security risk of MobiShare. In 2016, Liu et al. \cite{liu2016n} provided a system called BMobishare. Recently, Li et al. \cite{DBLP:journals/sj/LiYLCHW17} proposed a more secure location-sharing scheme. The aforementioned   systems support two kinds of queries, i.e., friends' queries and strangers' queries, and also satisfy access control policy. Aside from all listed above, \cite{du2009transactions,xiao2007survey,du2007effective} also provide efficient way for key management which bring support for cryptographic solutions.

However, these mechanisms are not perfect. Firstly, the threshold distance is a personal preference of each vehicle (to establish a social circle), but this is used as public information for location service entities in the system. When the threshold distance set by a vehicle is a special number, or the threshold distances set for different targets are in a special data group, the adversary can track the data or data group to identify vehicles. Secondly, threshold distance is used by a vehicle to determine with whom they are willing to share locations. Some schemes use broadcast encryption to share personal location information, which violates the distance-based access control policy. Finally, it is far from actual application requirements that all systems mentioned above use a single threshold distance for all friends. Vehicles may wish to set different threshold distances for different friends.

\textbf{Our contributions:}  Motivated by these issues, we propose a privacy-preserving location-sharing scheme in VSNs, namely PPLS. The contributions are described as follows.

\begin{enumerate}[(1)]

\item
In previous research, a vehicle can only set a single threshold distance for all friends. However, this setting does not meet the actual needs. To improve the practicability of the system, our scheme allows vehicles to set different threshold distances for different friends. In our scheme, vehicles can use a more flexible strategy to achieve access control.  
\item
Since existing works do not consider the privacy of the threshold distance, an adversary can easily collect threshold distances to get more personal information of vehicles. To overcome this defect,  we propose a new secure distance comparison protocol to execute encrypted distance comparison and prevent  location servers from determining this sensitive data. 
\item
Based on the proposed secure distance comparison protocol, we propose the  PPLS scheme. In PPLS, vehicles are allowed to set different threshold distances for different friends, and  broadcast encryption is not used, while diverse queries are used for information retrieval.

\end{enumerate}
     
This paper is organized in the following sections. In Section 2, we provide the system models and design goals. In Section 3, we present the building blocks including the proposed secure distance comparison protocol. Section 4 introduces the PPLS scheme and Section 5 gives its security analysis. In Section 6, performance analysis is provided. Finally, we draw a  conclusion in Section 7.

\section{System Models and Design Goals}
This section presents the formal system architecture, system work flows, and the threat model for location privacy. We also identify and list the security goals for the proposed scheme.

\subsection{System Architecture}
\begin{figure}[thb!]
\centering
\includegraphics[width=0.85\linewidth]{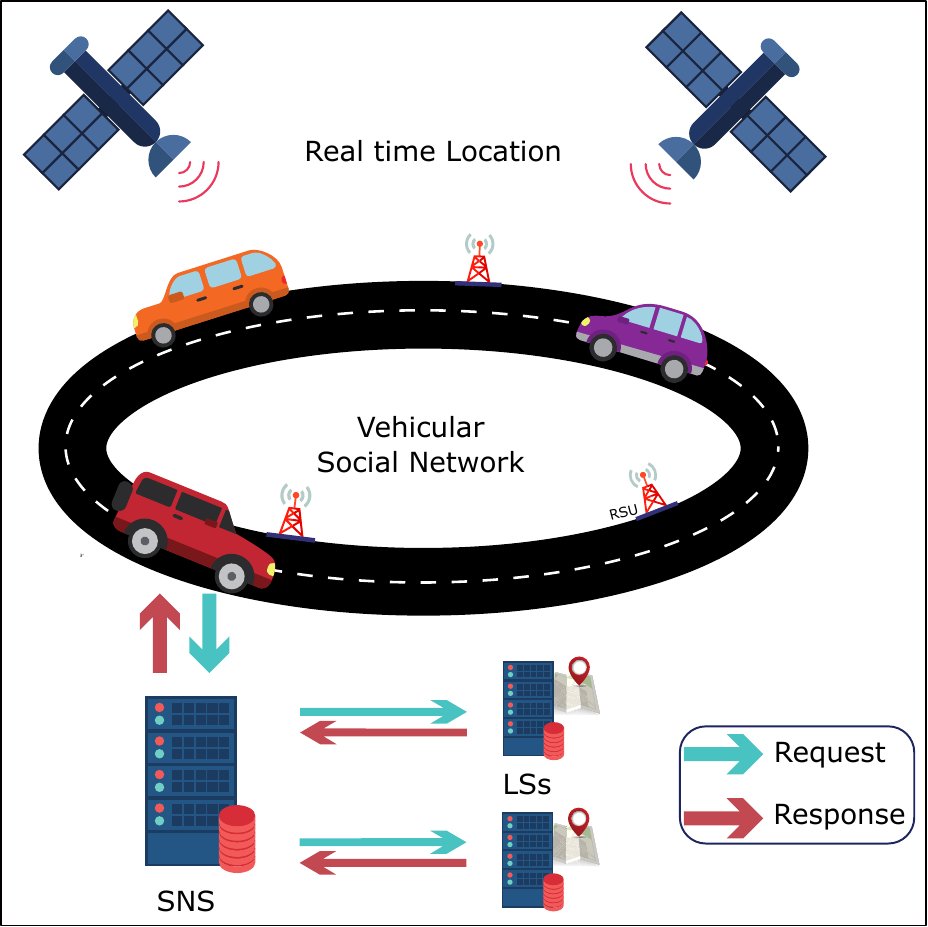}
\caption{System architecture }
\label{jiegou}
\end{figure}

The system architecture is depicted in Figure~\ref{jiegou} where four main entities interact with each other.

\paragraph{Vehicles}
 The vehicles of VSNs, can communicate with roadside units (RSUs) directly. They can get their own locations from GPS and request for locations of specific friends, nearby friends and strangers.
 
 \paragraph{RSUs}  After    RSUs receive requests from vehicles,  they  forward  them towards the social network server, then return the received   responses  to vehicles.

\paragraph{Social network server (SNS)}
 SNS is responsible for managing vehicles' social relationships, such as profiles and friend lists. SNS can communicate with  RSUs and location servers directly.	
\paragraph{Location servers (LSs)}
These servers primarily manage vehicles' location information. They calculate location distances and related tasks of finding vehicles within a certain area, which are assigned by SNS. LSs communicate with SNS directly, but different LSs are not allowed to cooperate with each other for information exchange.

\textbf{Constraints:} In our system, SNS should not be aware of the vehicle locations. Moreover, LSs are not aware about vehicles' identity-related information. Vehicles may submit three types of queries: 1) request for particular friends' locations, 2) request for nearby friends' locations, and 3) request for nearby strangers' locations.

\subsection{System Workflows}
In light of the proposed architecture, five main workflows are defined.

\begin{enumerate}[(1)]
\item
Vehicles must initially register with SNS for location based service. The registration process requires submitting personal identification information and make effective proof of authenticity. Moreover, vehicles must also define their access control policies. SNS maintains a database and processes vehicles' personal information. SNS registers all vehicles with the LS, using pseudo-identities and initial location information. 
\item
When arriving at a new place or after a specified time period, vehicles need to update their information. In this regard, SNS maintains the new relationships and threshold distances of vehicles, whereas LSs maintain the new location information.
\item
When a vehicle intends to obtain the location of a friend, they submit a query for that particular vehicle. If the requester meets the access control policies of their friends, they can obtain the location information.
\item
When a vehicle intends to obtain nearby friends current location information, they submit a query for friends within certain distance. If the vehicle meets the access control policies of these required friends, they can get the desired information.  
\item
In case of a vehicle requiring nearby stranger's current location, they submit a query for strangers within specific distance. If the vehicle meets the access control policy of strangers (within distance), they can get the locations of these strangers.  

\end{enumerate}

\subsection{Threat Model}
Out of the listed entities (i.e. vehicles, SNS, \& LSs), vehicles are considered to be dishonest. This means that they may try to access the server they do not have the permission to access, and find the location of a target vehicle. Moreover, we assume that SNS and LSs are honest but curious, i.e., they will follow the scheme formally, but try to obtain as much sensitive information as possible. For example, SNS may want to find the location of vehicles, and LSs may want to obtain sensitive information of vehicles. We suppose that SNS and LSs may be compromised by an adversary, but not at the same time. This means that SNS and LSs will not collude with each other. The assumption is reasonable since it is extremely difficult for an adversary to control the two servers at the same time.
 
\subsection{Security Goals}
Using the defined threat model as guiding principle, the security goals for location-sharing system are defined as below:
\begin{enumerate}[(1)]
\item
The system should protect vehicles' location information from SNS and other unauthorized vehicles. Vehicles' locations cannot be leaked to friends or strangers who do not satisfy the predefined access policy.
\item
SNS provides social relationships related service and should not be able to determine (directly or indirectly) the vehicle locations. 
\item
Location servers provide location-based services and should not know vehicles' social network information and/or identity information.
\end{enumerate}

\section{Building Blocks}
The main challenge to solve is to implement location-based services while preserving vehicles' privacy. In the proposed Privacy Preserving Location Sharing (PPLS) scheme, the vehicle sets threshold distances for different friends \& strangers, and the threshold values may vary  with different targets. It is important to note that, these values may indicate personal emotion tendency towards different targets, and location service providers can collect this data to infer such personal information. Therefore, the threshold distance should be kept private in addition to actual location. To solve this problem,  we propose a secure distance comparison protocol based on Paillier encryption. The scheme also makes use of RSA encryption, which is elaborated in a nutshell for comparative understanding.

\subsection{RSA Encryption}
RSA encryption is a widely used public-key cryptosystem for secure data transmission, where a public and private key pair is used for encryption and decryption. The process is summarized as:

Choose two large prime numbers $p$ and $q$, compute $n = pq$. Select random integer $e$ such that $1<e<\lambda(n)$ and $gcd(e,\lambda(n))=1$, where $\lambda(n)=(p-1)(q-1)$, and $gcd$ is the greatest common divisor. Compute $d=e^{-1}(\bmod(\lambda(n)))$. The public key is $(n,e)$ and the private key is $(p,q,d)$.
\paragraph{Encryption}
Assume that $M$ is a message to encrypt. First, turn $M$(un-padded plain text) into an integer $m$(padded plain text) by padding scheme. The ciphertext is $c=m^e(\bmod n)$.
\paragraph{Decryption}
Let $c$ be the ciphertext to decrypt, $m$ can be recovered by computing $c^d=(m^e)^d=m(mod n)$. The plain text message $M$ can be recovered by reversing the padding scheme.

\subsection{Paillier encryption}
Paillier public-key cryptosystem is a classical homomorphic semantically secure public-key cryptosystem, and is used in proposed secure distance comparison protocol. This section outlines the basic technique of Paillier public-key cryptosystem.

Choose two large prime numbers $p$ and $q$, and compute $n = pq$. Select random integer $g$, $g \in Z_{{n^2}}^*$, ensure $gcd(L(g^{\lambda}\bmod n^{2}),n)=1$, where 
$L(x) = \frac{{x - 1}}{n}$, $\lambda=lcm(p-1,q-1)$, and $lcm$ is the lowest common multiple. The public key is $(n,g)$ and the private key is $(p,q)$.
\paragraph{Encryption}

Assume that $m$ is a message to be encrypted where $0\le m\le n$. Select random $r<n$, then the ciphertext is $c = {g^m}\cdot {r^n}\bmod {n^2}$.
\paragraph{Decryption}

Let $c$ be the ciphertext to decrypt, where $c \in Z_{{n^2}}^*$, the plain text message is $m = \frac{{L\left( {{c^\lambda }\bmod {n^2}} \right)}}{{L\left( {{g^\lambda }\bmod {n^2}} \right)}}\bmod n$.

Paillier public-key cryptosystem has the following properties.
\paragraph{Homomorphic addition of plain texts}
We can give the value of $E({m_1}+{m_2})$ through $E({m_1})$ and $E({m_2})$ without knowing ${m_1}$ and ${m_2}$.

$D(E({m_1},{r_1}) \cdot E({m_2},{r_2})\bmod {n^2}) = {m_1} + {m_2}\bmod n$
\paragraph{Homomorphic multiplication of plain texts}
We can give the value of $E({m_1}{m_2})$ through $E({m_1})$ and ${m_2}$ without knowing ${m_1}$. 

$D(E{({m_1},{r_1})^{{m_2}}}\bmod {n^2}) = {m_1}{m_2}\bmod n$

\subsection{Secure Distance Comparison Protocol}
In our system, LSs need to compare the distance between two vehicles with the corresponding threshold distance to effectively provide services. To preserve vehicles' privacy, we propose a secure distance comparison protocol (as shown in Protocol~\ref{alg:compare}) based on \cite{lipmaa2003verifiable} and \cite{golle2006private}.
Let $d_{threshold}$ be threshold distance,  $g$ be a generator  of a cyclic group ${M}$, and $d_{actual}$ be the actual distance. We set $d_{threshold}$ and $d_{actual}$ as integers. $G$ is a key generation algorithm. $PE$ is the Paillier encryption algorithm, and $PD$ is the Paillier decryption algorithm. $R$ is the space of random coins. $S$ is a probabilistic polynomial time algorithm with $S({1^k},PK) \subset Z$, and $k$ is the security parameter. 

\floatname{algorithm}{Protocol}
\begin{algorithm}[h]
\caption{Security distance comparison protocol}
\label{alg:compare}
\begin{algorithmic}[1]
\Require
    Threshold distance $d_{threshold}$; Actual distance $d_{actual}$
\Ensure
    $d_{threshold}>d_{actual}$ as TRUE or FALSE
\State
    SNS generates the key pair $(sk_m,pk_m)\leftarrow {G}({1^k})$ and a random vale $r\leftarrow{R}$. Let $c\leftarrow{PE_{pk_m}(d_{threshold}g;r)}$. SNS sends $(pk_m,c)$ to LS;
\State
    LS  generates  random $s\leftarrow{S}$, $r'\leftarrow{R}$, computes
 \[ \begin{split}
c' \leftarrow &{(c \cdot {PE_{pk_m}}( - (d_{actual} + i)g;0))^s} \cdot {PE_{pk_m}}(0;r') \\=& {({PE_{pk_m}}(d_{threshold}g;r) \cdot {PE_{pk_m}}( - (d_{actual} + i)g;0))^s} \cdot {PE_{pk_m}}(0;r')\\=&{PE_{pk_m}}(s(d_{threshold} - (d_{actual} + i))g;{r^s} \circ r')
\end{split} \]  for $i=1,2,\ldots,n-1$,  LS computes $c'$ and send $c'$ to SNS;
\State
    SNS outputs $d_{threshold}>d_{actual}$ as TRUE, iff $PD_{pk_m}(c')=0$ is found. Otherwise output FALSE.
\end{algorithmic}
\end{algorithm}

\section{Privacy-preserving Location-sharing (PPLS) Scheme}
In order to preserve the vehicles' location and social network privacy, the scheme utilizes encryption keys generated by different system entities. The details of each step are given below, and Table~\ref{notations} lists the notations used in them.

\begin{table}[!h]
\begin{center}
\begin{tabular}{@{}ll@{}}
\hline
Symbol  &Description  \\
\hline
$ID$      &A vehicle's social network identifier\\
$PID$     &A vehicle's pseudo-identifier\\
SNS      &Social network server\\
LSs     &Location servers\\
$df$    &Threshold distance for a friend\\
$ds$    &Threshold distance for strangers\\
$(pk_u,sk_u)$   &A vehicle's public-private key pair\\
$(pk_m,sk_m)$   &SNS's public-private key pair\\
$(pk_s,sk_s)$   &LS's public-private key pair\\
$tl$     & The time length for LS  to save a record\\  
$ts$     & Time stamp\\
$t$      &Vehicles' location update cycle\\
$(x,y)$  &Location of a vehicle\\
$dis(u_i,u_j)$   &Distance between $u_i$ and $u_j$\\
$PE$       &Paillier encryption algorithm\\
$PD$       &Paillier decryption algorithm\\
\hline
\end{tabular}
\caption{Summary of notations}
\label{notations}
\end{center}
\end{table}	

\subparagraph{Initialization:}
Each vehicle has their identifier $ID$ and a public-private key pair $(pk_u,sk_u)$ which can later be updated. Assume SNS and LSs serve a designated area, and vehicles' group is represented as $U = \left\{ {{u_1},{u_2}, \ldots ,{u_z}} \right\}$. LS  has a public-private key pair $(pk_s,sk_s)$ and all vehicles know LS's $pk_s$.
\subparagraph{Registration:}
When a vehicle $u_i$ with an identifier $ID$ intends to use the system's services, they need to register with the SNS first. Registration is in the form of $(ID,{C_{pk_s}}(x_i,y_i), {C_{pk_s}}(pk_u), Flist,(df_{i,1},df_{i,2},...,ds), ts, Sig(ID,ts))$, where ${C_{pk_s}}(x_i,y_i)$ and ${C_{pk_s}}(pk_u)$ are $u_i$'s location \& public key (respectively) encrypted by LS's public key, $Flist$ is $u_i$'s friend list, $df_{i,1}$ is $u_i$'s threshold distance for friend $u_1$ within which they are willing to share location with $u_1$, $ds$ is the threshold distance for strangers with  which $u_i$ is willing to reveal its location to strangers, $ts$ is a time stamp, and $Sig\left( {ID,ts} \right)$ is a signature generated on $ts$. SNS holds a database to save vehicles' threshold distances. 



SNS confirms the request. If the signature is valid, SNS generates a registration request to LS. The request is in the form of $\left( PID,{C_{pk_s}}\left( {x,y} \right),{C_{pk_s}}\left( {pk_u} \right),tl\right)$, in which $PID$ is $u_i$'s pseudo-identity generated by $AES\left( {ID,rt} \right)$ and $rt$ is a random value. $tl$ is the time limit for which the record will be held. LSs can timely remove the expired data and reduce storage overhead. The value of $tl$ should be set slightly larger than the update cycle.
\subparagraph{Update:} 
For each time period $t$, vehicles need to update their information. Similar to the registration content, each vehicle sends a message to SNS in the form of $(ID,{C_{pk_s}}\left( {x,y} \right),{C_{pk_s}}\left( {pk_u} \right),Flist,(df_{i,1},df_{i,2},...,ds),ts,Sig(ID,ts))$, where ${C_{pk_s}}(x,y)$, $Flist$ and $(df_{i,1},df_{i,2},...,ds)$ represent vehicle's new location  encrypted by LS's public key, new friendship, and new threshold distances. Without updating $pk_u$, the adversary can associate the vehicle's $PID$s by tracing $pk_u$. If the signature is valid, SNS sends $\left(PID,{C_{pk_s}}\left( {x,y} \right),{C_{pk_s}}\left( {pk_u} \right),tl \right)$ to LSs. LSs save related information in their database.

\subparagraph{Request for particular friends:}
If a vehicle $u_i$ with an identifier $ID$ wants to obtain the location(s) of their friend(s) $\left( {{f_1},{f_2}, \ldots ,{f_M}} \right)$, $u_i$ submits a query for friends' locations in the form of $( ID,{C_{pk_s}}( {x_i,y_i}) $,$pf$,$ ({{f_1},{f_2}, \ldots, {f_M}}) )$ to SNS, where $pf$ represents the request type. To handle this request, SNS first recovers the pseudo-identity $P_{ID}=(PID_1,PID_2,\ldots,PID_M)$ corresponding to $(f_1,f_2,\ldots,f_M)$. Then, SNS randomly divides $P_{ID}$ into $Q$ subsets $P_{ID}^1,P_{ID}^2,\ldots,P_{ID}^Q$ with different sizes, satisfying $P_{ID}=P_{ID}^1 \cup P_{ID}^2 \cup\ldots\cup P_{ID}^Q$, to prevent the adversary from knowing $u_i$'s friend relationships. For $P_{ID}^j=(PID_1,PID_2,\ldots,PID_N)$, SNS computes $(c_{1,i},c_{2,i},\ldots,c_{N,i})=(PE_{pk_m}(df_{1,i}g;r_1),PE_{pk_m}(df_{2,i}g;r_2),\ldots,$\\$PE_{pk_m}(df_{N,i}g,r_N))$, and sends $(PID,{C_{pk_s}}\left( {x_i,y_i} \right)$,$pf$,$P_{ID}^j,(c_{1,i},c_{2,i},\ldots,c_{N,i}),pk_m)$ to $LS_j$, where $LS_j$ is the $j$th location server in LSs. After receiving the request, $LS_j$ performs the following steps:\\
\begin{enumerate}[(1)]
\item
Decrypt $C_{pk_s}(x_i,y_i)$ to get $u_i$'s current location $(x_i,y_i)$.\\
\item
Calculate the distances between $u_i$ and its friends, and save as $(d_1,d_2,\ldots,d_N)$\\$=\left(dis({u_i},{PID_1}),dis(u_i,PID_2), \ldots ,dis(u_i,PID_N)\right)$.\\
\item
Choose parameters $s$ and $r'$. For $c_{1,i}$, calculate \[{c'_{1,i}} = \left( {{{\left( {{c_{1,i}} \cdot P{E_{p{k_m}}}\left( { - \left( {{d_1} + p} \right)g;0} \right)} \right)}^s} \cdot P{E_{p{k_m}}}\left( {0;r'} \right)} \right).\]
 
Let $p=1,2,\ldots,n-1$ and send corresponding ${c'_{1,i}}$ to SNS.\\
\end{enumerate}
If and only if there exists $p$ which makes $PD_{sk_m}(c'_{1,i})=0$, then $d_1<df_{1,i}$ and $u_i$ satisfies $PID_1$'s access control policy, otherwise $u_i$ does not satisfy the policy. SNS finds all $u_i$'s friends for whom $u_i$ satisfies their access control policies. Then $LS_j$ sends those friends' encrypted locations to SNS.
After collecting all results returned by LSs, SNS sends $u_i$ the ciphertexts. $u_i$ decrypts the ciphertexts and gets their requested friend's location.

\subparagraph{Request for friends within specific distance:}
If a vehicle $u_i$ with identifier $ID$ wants to find friends' locations within a certain distance, then a query for friends' locations is submitted in the form of $(ID,C_{pk_s}(x_i,y_i)$,$f$,$l)$ to SNS, where $f$ indicates the type of request. Similar to request for particular friends' locations, after grouping friends randomly, SNS sends $(PID,C_{pk_s}(x_i,y_i)$,$\\$$f$,$P_{ID}^j,(c_{1,i},c_{2,i},\ldots,c_{N,i}),pk_m,l)$ to $LS_j$. When receiving the request, $LS_j$ performs the following steps:\\
\begin{enumerate}[(1)]
\item
Decrypt $C_{pk_s}(x_i,y_i)$ to get $u_i$'s current location $(x_i,y_i)$.\\
\item
Calculate the distances between $u_i$ and all of their friends, and save as $(d_1,d_2,\ldots,d_N)=\left(dis({u_i},{PID_1}),dis(u_i,PID_2), \ldots ,dis(u_i,PID_N)\right)$.\\
\item
Choose parameters $s$ and $r'$. For $c_{1,i}$, calculate \[{c'_{1,i}} = \left( {{{\left( {{c_{1,i}} \cdot P{E_{p{k_m}}}\left( { - \left( {{d_1} + p} \right)g;0} \right)} \right)}^s} \cdot P{E_{p{k_m}}}\left( {0;r'} \right)} \right).\] Let $p=1,2,\ldots,n-1$ and send corresponding ${c'_{1,i}}$ to SNS.
\end{enumerate}
If and only if there exists $p$ which makes $PD_{sk_m}(c'_{1,i})=0$, then $d_1<df_{1,i}$ and $u_i$ satisfies $PID_1$'s access control policy. Furthermore, if $d_1<l$, $f_1$'s location will be returned. SNS finds all these friends and gets their encrypted locations from $LS_j$. After collecting  the results returned by all LSs, SNS sends the final response to $u_i$, which decrypts the ciphertext with their own private key $sk_u$ and gets the friends' locations.

\subparagraph{Request for strangers within specific distance:}
If a vehicle $u_i$ wants to find location of stranger(s) who are within $l$ distance from them, then $u_i$ submits a strangers' locations query $(ID,C_{pk_s}(x_i,y_i)$,$s$,$l)$ to SNS. Here $s$ is the request type. Since there are too many unfamiliar vehicles around $u_i$, SNS sends LSs a query $(PID$,$all$,$l)$ first. LSs find all vehicles within $l$ distance away from $u_i$ and feed back the result. Then, SNS eliminates $u_i$'s friends randomly, and sends $(PID,C_{pk_s}(x_i,y_i),s,P_{ID}^j,(c_{1,i},c_{2,i},\ldots, c_{N,i}),pk_m)$ to $LS_j$. Assuming a stranger $u_2$ is within $l$ distance away from $u_i$. $u_2$'s location is $(x_2,y_2)$ and $u_2$'s threshold distance for strangers is $ds_2$. If and only if $dis(u_i,u_2)<ds_2$, $LS_j$ returns $u_2$'s encrypted location to SNS. SNS then sends the final result to $u_i$.

\section{Security Analysis}
The security analysis is provided based on the threat model and security goals. In PPLS, we assume that SNS and LSs. Hence, they do not collude with each other, and are not compromised by the adversary at the same time.

\paragraph{Access control}
PPLS allows vehicles to set different threshold distances for different targets. Since SNS and LSs are assumed to be honest but curious, they will follow the protocol formally. That means, only the vehicles who satisfy the access policy can receive the location information and identity information of friends/strangers. 

\paragraph{Identity privacy}
In PPLS, LSs should not have any knowledge of vehicles' identity-related information. Pseudo-identity is used when vehicles send update messages or queries. Thus, anonymity is achieved. Though threshold distances may leak identity information (indirectly) of vehicles to the adversary, homomorphic encryption is used to encrypt the sensitive data. Thus, vehicles' identity privacy is well preserved.
\paragraph{Location privacy}
SNS may collude with dishonest vehicles and attempt to obtain the location information of a particular vehicle illegally. When receiving the registration/update messages from vehicles or receiving the responses from LSs, SNS has the chances to access vehicles' locations. PPLS encrypts vehicles' locations using asymmetric encryption, which protects location information from SNS. 

\paragraph{Social network privacy}
The privacy of the social network is preserved by two approaches, which are described as follows.  
\begin{enumerate}[(1)]
\item
When a vehicle requests for particular friends or friends/strangers within specific distances, SNS will divide the friends/strangers into random subsets and send these sets to different LSs. These subsets have different sizes and will be sent to LSs randomly. Furthermore, dummy vehicles can be added into the original set. As a result, each LS can only get part of the friend list with dummy vehicles. Since we assume  that LS will not collude with each other,  LSs are prevented  from knowing vehicles' social networks.
\item
For each time period $t$, vehicles need to update their information. During this phase, SNS assigns each vehicle a new pseudo-identifier, which is different from the original one. As a result, after the time period $t$, for different queries from the same vehicle, the vehicle's pseudo-identifier and its friends' pseudo-identifiers become different. Therefore, it is impossible for LSs to determine the information of vehicles' social networks. 
\end{enumerate}

\section{Experimental Evaluation}
The proposed PPLS scheme uses a number of encryption and decryption steps. To evaluate the real time performance, we have conducted a number of experiments.

\subsection{Implementation}
In our system, three cryptography schemes are implemented: digital signature, asymmetric encryption, and homomorphic encryption. We use RSA \cite{DBLP:conf/eurocrypt/BellareR96} with 1024-bit key size for data encryption, RSA PKCS1-v1-5 for signature, and Paillier with 1024-bit key size for homomorphic encryption. Our simulation is implemented on an Intel Xeon E3-1230v3 running at 3.4 GHz with 8 GB 2133 GHz memory. We use Python 3.5.0 to implement the proposed algorithms. Some PyPI packages are used in our cryptography schemes: $pycrypto$ for signature, asymmetric encryption and $phe$ for Paillier encryption.

In our experiments, vehicles can use many effective techniques to obtain locations, such as GPS. We assume that the threshold distance can set as ${10,20,\ldots,100}$ meters with steps of 10 meters or ${100,200,\ldots,1000}$ meters with steps of 100 meters. For friends, vehicles may consider choosing a smaller value as the threshold distance. For strangers, vehicles may choose a larger value as the threshold distance.

\subsection{Evaluation}
As the RSA signing technology used in the registration and update phases can be replaced by any other signing algorithms, we do not analyze the registration and updating phase. 

\begin{figure}[thb!]
\centering
\includegraphics[width=0.8\linewidth]{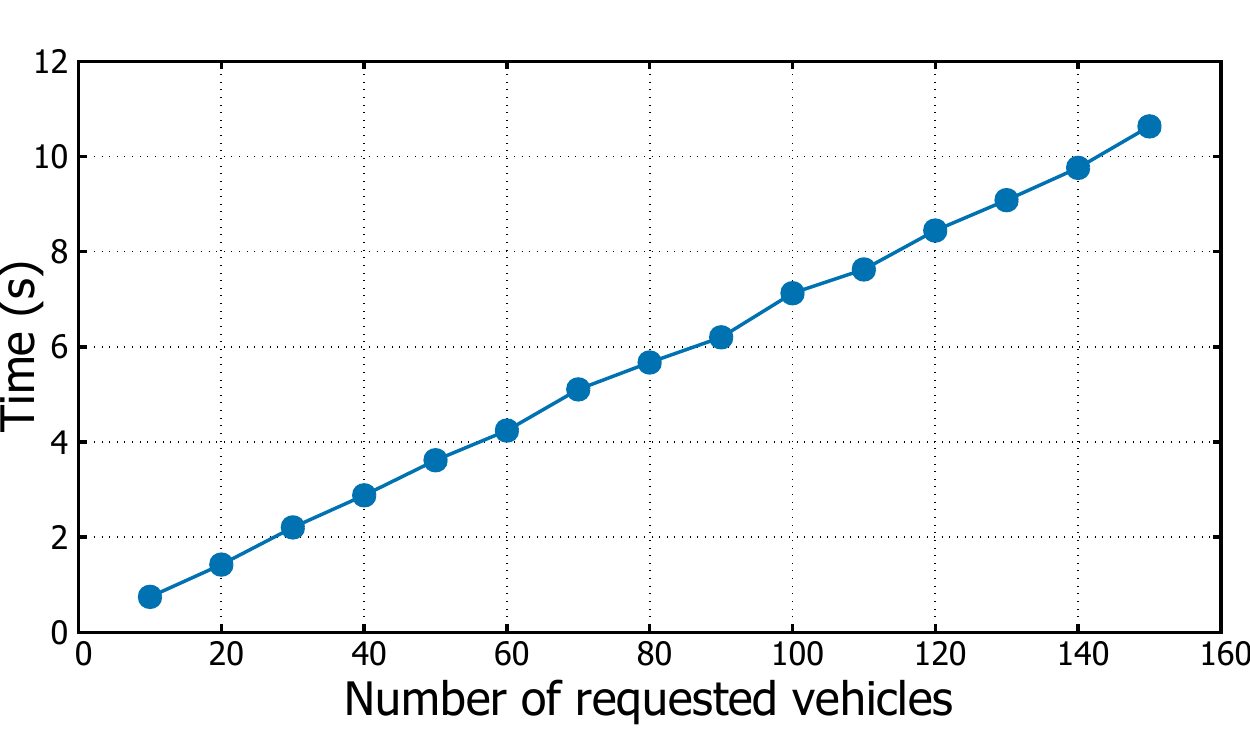}
\caption{Entire request process}
\label{shiyan1}
\end{figure}

\begin{figure}[thb!]
\centering
\includegraphics[width=0.8\linewidth]{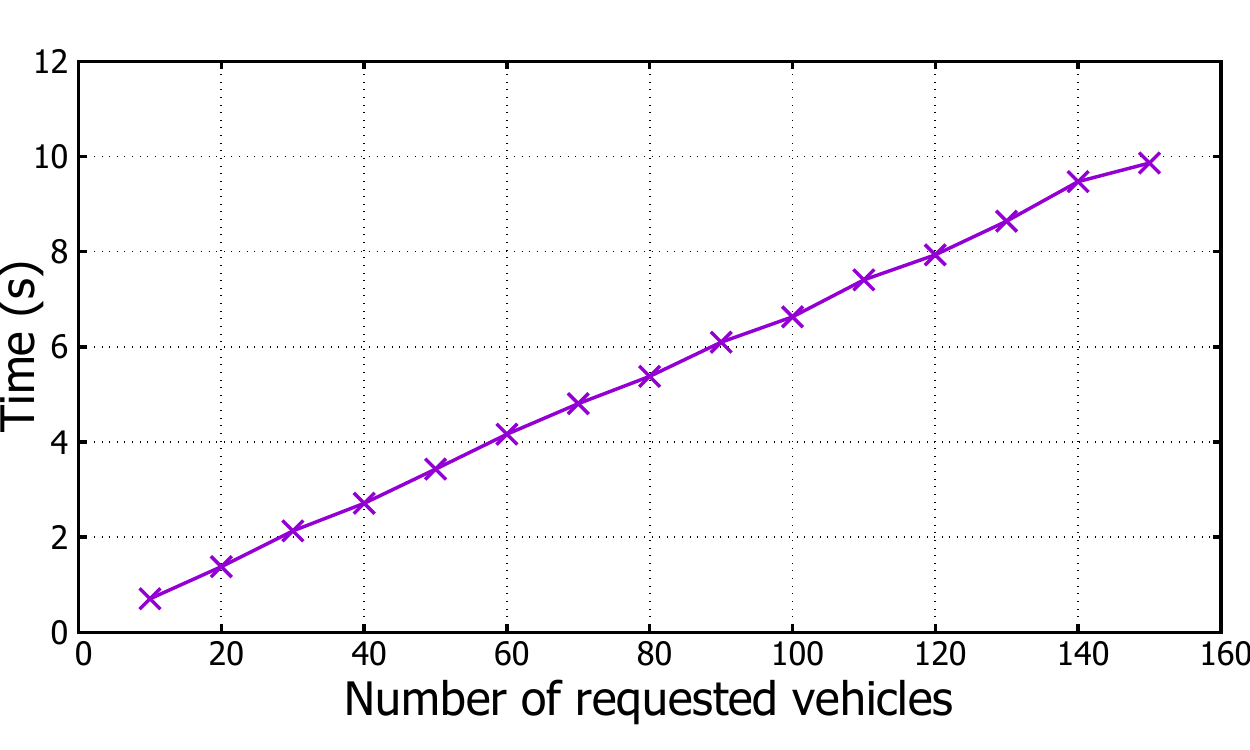}
\caption{Secure distance comparison protocol process}
\label{shiyan2}
\end{figure}

The response time of the system to request for particular friends is related to the number of friends the vehicle requests. The response time to request for friends or strangers within specific distance is related to the size of the request area and the vehicle density within the scope. In essence, this parameter is also based on number of vehicles requested. Therefore, we observe the time spent for entire request process and the time spent for secure distance comparison protocol against different number of requested vehicles. We conduct each experiment 10 times and calculate the average values. The results are  shown in Figure \ref{shiyan1}  and   Figure \ref{shiyan2}, respectively.

It can be observed from the results that the time spent on the request process increases approximately linearly with the number of vehicles requested, about 0.75 seconds per 10 individuals. The secure distance comparison protocol execution time also increases approximately linearly with the number of vehicles requested, about every 10 individuals with 0.7 seconds. The time spent in  implementing the secure distance comparison protocol takes up a large percentage of the system's time (in order to generate a response). The protocol time-consumption is mainly focused on determining the size relationship between the actual distance and the threshold distance, the traversal encryption of the actual distance in LS and the response decryption in SNS.

\section{Related Works}
In recent years, mobile computing has changed the future of communications and sevices \cite{Zhang2005,Xiao2007}, and accordingly promotes the rapid development of vehicular networks, VSNs have experienced an explosive development. Since a vehicle's location is important information used in VSNs, the issue of protecting vehicles' location privacy has received considerable attention. Until now, many studies on location privacy protection \cite{ju2015location,rao2015novel} have been done, such as location anonymity, information hiding \cite{DBLP:journals/ijcomsys/Das17} and so on. Location anonymity is an effective technique for location privacy protection and there are two types of methods to achieve it: 
1)~$K$-anonymity: The fundamental premise is to mix the real user's location information into $k-1$ other anonymous users' location information, which confuses the adversary. This approach is proposed in \cite{DBLP:journals/ijufks/Sweene02} by Sweeney in 2002, and then Gruteser et al. used it for location privacy protection in \cite{DBLP:conf/mobisys/GruteserG03}. Kido et al. extended $K$-anonymity, and  introduced the concept of virtual location \cite{DBLP:conf/icde/KidoYS05}. 
2)~Location encryption: The main idea of Location encryption  is to   encrypt the users' location information with some encryption algorithms, such as the algorithm proposed by Khoshgozaran et al.  \cite{DBLP:conf/ssd/KhoshgozaranS07} using Hilbert curves to encrypt the original location.

By combing the  aforementioned methods, a series of  research works have been proposed. In 2007, SmokeScreen \cite{DBLP:conf/mobisys/CoxDM07} proposed a scheme to   protect users' location privacy and provide location-sharing services for users. Subsequently, Wei et al. proposed MobiShare \cite{Wei2012MobiShare}, which supports users sharing location information flexibly. In MobiShare, social network server and location server store users' profiles and location information separately. Hence, neither of the two severs know the complete information of the users. However, this scheme cannot protect users' social network topologies. Later, based on MobiShare, several mechanisms were proposed, such as  N-MobiShare \cite{shen2016efficient,DBLP:conf/3pgcic/LiuLCLJ13}, MobiShare+ \cite{li2014mobishare}, and B-MobiShare \cite{shen2016efficient}. In N-MobiShare, cellular tower was not treated as a core component of the system. Social network server took cellular tower's task and forwarded users' requests to location server. N-MobiShare used broadcast encryption to share off-line keys to users' friends. Although N-Mobishare has a simpler structure than MobiShare, it did not solve the problem which MobiShare suffered. That is, the location server can still get users' social network topologies in the query phase. 
Inspired by Wei et al.'s solution, Li et al. found that in MobiShare the pseudo-identity of the querying user can be known by LSs in the friend's query. Hence, they proposed an improved mechanism named MobiShare+ \cite{li2014mobishare}. Besides dummy locations and identities, this mechanism employed dummy queries. It applied a private set intersection protocol to prevent individual information leaked between the social network sever and the location server. MobiShare+ overcomes the weakpoints of MobiShare and N-MobiShare. However, it incurred excessively long processing time. To solve this problem and improve the transmission efficiency, Shen et al. proposed B-MobiShare \cite{shen2016efficient}. Bloom Filter was used in this scheme to replace the private set intersection protocol in MobiShare+ and the time cost was reduced. However, B-MobiShare was less efficient than expected, the time cost was still high. In 2017, Li et al. proposed a system with enhanced privacy \cite{DBLP:journals/sj/LiYLCHW17}, using multiple location servers to prevent insider attack launched by the service providers.

However, all the above mechanisms do not treat the threshold distance as sensitive data, and work with a single threshold distance for users to set for all of their friends, which is unrealistic in real social networks. 

\section{Conclusion}
Privacy preservation of location sharing in VSNs is an important issue. In this article we propose PPLS, which protects vehicles' location privacy from SNS and preserves vehicles' social network privacy from LSs. The scheme allows vehicles to set different threshold distances for different friends, and to enjoy a more flexible access control policy. In order to implement this access control policy, a secure distance comparing protocol is presented. To permit vehicles sharing locations with friends, new queries are designed for particular friends. The security analysis shows that PPLS is secure under a comprehensive security model. Moreover, the experimental evaluation demonstrates  the efficiency of PPLS.

\section*{References}
\bibliography{ref}
\end{document}